 \documentclass[letterpaper,prb,tightenlines]{revtex4}
 
 \usepackage{graphicx}
 \usepackage{dcolumn}
 \usepackage{amsmath}
 \usepackage{hyperref }
\begin{document}
 \draft
 
 \bibliographystyle{prsty} 

 \title{ Polarization-dependent Intensity Ratios in Double Resonance Spectroscopy }
 
 \author{Kevin K. Lehmann}
\affiliation{
Departments of Chemistry and Physics, University of Virginia, Charlottesville VA, 22904-4319}
 \date{\today}

\begin{abstract}
Double Resonance is a powerful method spectroscopic method that provides unambiguous assignment of the rigorous quantum numbers of one state of a transition.   
However, there is often ambiguity as to the branch ($\Delta J$) of the transition.  The dependence of the intensity of the double resonance signal on the relative polarization
of pump and probe radiation can be used to resolve this ambiguity and has used for this in the past.  However, the published theoretical predictions for this ratio are based upon
a weak (i.e. non-saturating) field approximation.  In this paper, we present theoretical predictions for these intensity ratios for cases where the pump field is strongly saturating, 
in the two limits of transitions dominated by homogeneous and inhomogeneous broadening.  While saturation, as can be expected, reduces the magnitude of the polarization
effect (driving the intensity ratio closer to unity), polarization anisotropy remains even with a strongly saturating probe field in most cases.   For the case of an inhomogeneous 
broadened line, as when Doppler broaden linewidth dominates over even the power broadened homogeneous line width, a large fraction of the low pump power anisotropy
remains.  Results are presented for both the case of linear and circular pump and probe field polarizations. The present predictions are compared with experimental measurements
on CH$_4$ ground state $\rightarrow \nu_3 \rightarrow 3\nu_3$ transitions recently reported by de Oliveira et al and found to be in better agreement than the weak field
predictions.
\end{abstract}

 \maketitle

Double resonance (DR) has long been one of the most powerful methods in the spectroscopist's toolkit.\cite{carroll66}  This is an intrinsically nonlinear spectroscopy using two coherent light sources, at least one (the pump) that creates a nonequilibrium population distribution in a sample, and another (the probe) that measures an absorption, emission, scattering, or action spectrum of the resulting nonequilibrium sample.  There are thus three states linked by two transitions.  We neglect in this paper ``4-level'' DR transitions\cite{Pursell90, Mandal93} where a collision transfers population from a state of one transition to a state of the other.   DR results in greatly simplified spectra \cite{Teets76, Coy86a, Coy86b, Lehmann88, Coy89, Lehmann86, Gambogi94, Gambogi96, Kabir03, Zhao04}  and allows the unambiguous assignment of the starting state of probe transitions, which is often required when spectra are highly perturbed and do not follow regular patterns due to the breakdown of the separation of degrees of freedom, such as vibration and rotation.\cite{Park15}   It is also useful for observation of spectra from states without significant thermal population \cite{Martinez04, Settersten03, Foltynowicz21a, Foltynowicz21b}  and cases where homogeneous or inhomogeneous broadening results in substantial overlap of individual transitions, which can result, for example, in a broad rotational contour without resolvable features.\cite{Henck95a, Henck95b, Callegari97, Callegari00, Callegari03}  DR allows the selective population of states that have negligible thermal populations under available experimental conditions, allowing novel spectroscopic transitions to be observed. \cite{Barnum20, Yang20}  In many cases, at least some of the transitions from these states reach final states that are weak or forbidden from the thermally well-populated states due to symmetry or propensity selection rules.\cite{Chevalier87,Frye87, Kasahara94, Kasahara99, Srivastava20b, Dia21}  When DR is performed using narrow bandwidth lasers, one can largely eliminate inhomogeneous broadening, as the pump laser will produce a Bennet hole\cite{Elbel90} in the velocity distribution of the initial state and corresponding Bennet hill in the upper state of the pump transition.   Probe spectra will display sub-Doppler features whose widths are on the order of the homogeneous widths, which can be orders of magnitude below Doppler broadened widths. \cite{Kaminsky76, Wieman76}  Due to elastic collisions, the narrow feature will often ride on top of a Doppler broadened feature, but the former usually has a much higher peak strength.  DR, especially with a pulsed pump source and  continuous wav (CW) probe field, has been widely used to study elastic, reorientation, and inelastic collision rates and kernels.  \cite{Greene83, Coy92, Klaassen94, Coy95, Shin92, Soriano98, Abel99, Bayram06, Forthomme15}  Resonant 3-wave mixing, which is another form of DR spectroscopy, can be used to both measure enantiomeric excess of chiral molecules \cite{Patterson13a, Patterson13b, Lobsigner15, Lehmann18a} and to create enantiomeric excess in single rotational states. \cite{Eibenberger17}

  If the common level in the DR scheme is the lowest energy level (called V-type DR), the DR signal will be a narrow depletion of the background Doppler Broadened probe transition.  If the common level is not the lowest energy state, the pump will create a narrow absorption or emission depending upon whether the final state is higher or lower in energy than the pumped (intermediate) state.  These are known as ladder- and $\Lambda$-type DR respectively.  If there is a negligible thermal population in the intermediate state, the probe transitions will be a new narrow transition.   The very narrow width of DR transitions can be a drawback when one needs to sample the probe transition spectrum over a broad spectral range (say 30\,THz).  The time required to search such a spectral range will on the order of 10 times the detection time constant times the ratio of the scan range divided by the width of the probe DR transitions.   In such cases, detection of probe transitions of a few MHz or less width requires either very long search times or a very short time constant, which reduces the signal to noise ratio (SNR) of the probe spectrum.   The probe spectrum scan needs to be repeated for each pump transition studied.   The recent demonstration of DR using a stabilized frequency comb for the probe allows one to simultaneously sample the probe spectrum at the frequencies of tens of thousands of comb teeth has dramatically reduced this limitation. \cite{Nishiyama16, Foltynowicz21a, Foltynowicz21b}

Selection rules greatly reduced the final states that are observed in double resonance using a particular pump transition.   However, one often retains an ambiguity as to the total angular momentum quantum number $J$ of the final state of a probe transition due to the $\Delta J = 0, \pm 1$ selection rule for dipole transitions. The pump transition produces a nonequilibrium alignment of the angular momentum projection quantum number $M$ of the initial and final states, and this results in the probe absorption strength depending upon the relative polarization state of the pump and probe fields, which was first reported by Frankel and Steinfeld (1975).\cite{Frankel75}   By comparing the strength of the DR signal with parallel and perpendicular linear polarizations, one obtains a polarization ratio that can be used to assign the value of $\Delta J$ for the probe transition.   One sensitive implementation of DR is polarization spectroscopy which uses nearly crossed polarizers for the probe wave, placed before and after the sample.\cite{Li04, Steeves05}  The pump laser-induced birefringence of the sample results in a change in the transmission of the probe that is observed on a greatly reduced background intensity which will result in an increased signal-to-noise ratio if the probe field is dominated by technical intensity noise.   Another sensitive variation is polarization modulation where a change in probe transmission is produced by a polarization modulation of the pump field.\cite{Shin92, Soriano98} In the limit that the pump transition is not saturated, the predicted polarization dependences of the DR signals are easily derived from the dependence of pump and probe field transition intensity on $M$.   Expressions have been given in the literature \cite{Chevalier87, Jonas92, Ferber97, Kabir03, Chen08, Petrovic08}  though often not in a form most useful for DR spectroscopy, and almost always neglecting the effects of optical saturation.   Significantly, one optimizes the strength of DR transitions by working with sufficient pump power to have substantial saturation of the pump transitions as that produces a larger pump-induced disequilibrium of the sample yet quantitative discussion of these cases is largely absent.   One of the few exceptions was the study of Spano and Lehmann \cite{Spano92} who examined the case of polarization spectroscopy of a sample that is optically thick for the pump transition.  It was found that for excitation of a dipole transition with a pulse duration much shorter than the relaxation time, a strong pump pulse evolves into an area-preserving pulse similar to the self-induced transparency of a two-level system.\cite{McCall69}  This pulse produces an even greater fractional alignment of the sample than that produced by excitation with negligible saturation.   For DR with continuous wave pump fields, which produce a steady-state response of the sample, the analysis of Spano and Lehmann\cite{Spano92} is not applicable.   In this paper, we will present an analysis of the polarization dependence of DR signal strength for cases of a steady-state response of the sample.   First, the weak field case will be considered, which results in simple analytical expressions.  This is followed by results for a saturating pump field both for the cases of homogeneous and inhomogeneous broadened transitions.   

\section{ Polarization Dependence of  Pump Transitions}

Consider a DR signal that results from a pump transition between a pair of levels  $1$ and $2$ and a probe transition between levels $2$ and $3$, and label the total angular momentum quantum numbers for the three levels as $J_1, J_2$, and $J_3$ respectively.  Let the pump (probe) transition be driven by waves $a (b)$ with angular frequency and wavevector $\omega_{a,b}$ and $\vec{k}_{a,b}$ respectively.  
Each $J_2, M$ state will contribute to the DR signal proportional to the population change in that state produced by the pump laser, $\Delta \rho_{22}(M, \Delta \omega_{12})$ where $\Delta \omega_{12}$ is the detuning of the pump from resonance, times the absorption coefficient of the probe laser by that state, $S(M)$, both of which will be dependent on $M$ and the polarization directions of pump and probe fields respectively. Both pump and probe strengths depend upon the respective transition dipole moment matrix element for the respective transition which has the form  $\left< i, M | \vec{\mu} | j , M' \right> \cdot \hat{G} = \left< i | \mu_g | j \right>  \left< i, M | \phi_{gG} | j, M'\right>$.  where $g$ gives the direction of the transition dipole moment, $\mu$, between states $i$ and $j$ in the molecular frame, $\hat{G}$ gives the direction of the optical electric field, $E$ in the laboratory fixed frame, and  
$\left< i, M | \phi_{gG} | j, M' \right>$ is the direction cosine matrix element which is the matrix element of $\hat{g} \cdot \hat{G}$.  The 
transitions direction cosine matrix elements are given in Table~4.4 of \textit{Microwave Spectroscopy} by Townes and Schawlow\cite{Townes&Schawlow} and reproduced in Table~\ref{DirCosineElements} for completeness.  The direction cosine matrix elements consist of three factors but only the one that depends on $J$ and $M$ for each state, $\phi_G (J,M,J'.M')$, of a transition is needed for predicting the polarization dependence -- the other two factors are independent of $M$ and the polarization state of the radiation field.  

We begin by considering linear polarization for the pump and probe fields.  The total signals are independent of how we align the laboratory axes; we will take the $Z$ axis as along the polarization direction of the pump wave and $Y$ axis along the propagation direction of both pump and probe fields.  With this axis assignments, we use  $\phi_Z(J_1,M,J_2,M)$ for the pump matrix element and for the probe $\phi_Z(J_2,M,J_3,M)$ for parallel and $\phi_X(J_2,M, J_3,M\pm1)$ for perpendicular relative polarizations. 

Treating the pump transition as a two-level system for each $M$ value, the steady-state change in population in each $M$ state of level $2$ can be written in terms of the equilibrium population density difference between levels $1$ and $2$, $\rho_{11}^e - \rho_{22}^e$; the population and coherence relaxation rates of the pump transition, $\gamma_1, \gamma_2$; the pump Rabi frequency $\Omega_{12}(M) = \mu_{12}(M) E /\hbar $ where $E$ is the amplitude of the field driving the $1 \leftrightarrow 2$ pump transition; and the detuning from resonance, $\Delta \omega_{12} = \omega_a - \vec{k}_a \cdot \vec{v} - \omega_{12} $ with $\vec{v}$ the velocity of the absorber:\cite{Karplus48, Townes&Schawlow}
\begin{equation}
\Delta \rho_{22}(M, \Delta \omega_{12}) = \frac{( \rho_{11}^e - \rho_{22}^e )}{2 (2 J_2 +1)}    \frac{ (\Omega_{12}(M)^2 \gamma_2 /\gamma_1) }{\Delta \omega_{12}^2+\gamma_2 ^2\left(  1 +  (\Omega_{12}(M)^2/\gamma_1 \gamma_2)  \right) }   \label{delta_rho}
\end{equation}
This is a Lorentzian with half width half maximum (HWHM) of $\gamma_2 \sqrt{  1 +  (\Omega_{12}(M)^2/\gamma_1 \gamma_2)   }$.  
We will treat the multistate transition with different values of $M$ as
a sum of two-level systems with different transition dipole moments (hence the $M$  labels in Eq.~\ref{delta_rho}).  Schwendeman\cite{Schwendeman80,Schwendeman99} has pointed out
this is an approximation, but concluded it holds if one neglects pure $M$ changing collisions (elastic $J$-reorientation).  
We will assume that the probe transition is unsaturated and thus each $M$ value has an absorption strength $S(M)$ proportional to $\phi_Z(J_2,M,J_3,M)^2$ 
for $\parallel$ alignment and $\phi_X(J_2,M,J_3,M+1)^2$+$\phi_X(J_2,M,J_3,M-1)^2$ for $\perp$ alignment.

The total DR signal ratio for a given probe polarization, $G$, can be written as:
\begin{equation}
R_{\rm lin} =   \frac{I_{\parallel}}{I_{\perp}} =  \frac{  \sum_M  (\int  \Delta \rho_{22}(M, \Delta \omega_{12}) d\omega_{12})  \phi_Z(J_2,M,J_3,M)^2 }
{ \sum_M  (\int  \Delta \rho_{22}(M, \Delta \omega_{12}) d\omega_{12}) ( \phi_X(J_2,M,J_3,M+1)^2   +  \phi_X(J_2,M,J_3,M-1)^2 )  } 
\end{equation}
If the transition is homogeneously broadened, there is only a single value of $ \Delta \omega_{12}$ and thus not integral of this detuning.

\subsection{Unsaturated pump transitions}
In most CW gas-phase DR experiments, the pump Rabi frequencies, $\Omega_{12}(M)$ are far below the Doppler width of the pump transition,
in which case the pump will burn a Bennet hole in the velocity distribution of the lower energy state and a Bennet hill in the upper energy state.
If we assume that the pump transition is inhomogeneously  Doppler broadened with lineshape function $g_{\rm D}$ that has width $\Delta \omega_{\rm D} >> \gamma_1, \Omega_{12}(M)$, we can integrate Eq.~\ref{delta_rho} over the Doppler detuning to give an integrated steady-state population change for level $2$
\begin{equation}
\Delta \rho_{22}(M) =  \frac{\pi}{2 (2 J_2 +1)}  ( \rho_{11}^e - \rho_{22}^e )  g_{\rm D}(\omega - \omega_{12})  \frac{  (\Omega_{12}(M)^2 }{ \gamma_1 \sqrt{  1 +  (\Omega_{12}(M)^2/\gamma_1 \gamma_2) }}  \label{eq:drho2}
\end{equation}
In the limit of low saturation, $\Omega_{12}^2(M) << \gamma_1 \gamma_2$, get 
\begin{equation}
\Delta \rho_{22}(M) \rightarrow  \frac{\pi}{2 (2 J_2 +1)}  ( \rho_{11}^e - \rho_{22}^e )  g_{\rm D}(\omega_a - \omega_{12})  | \Omega_{12}(M) |^2  / \gamma_1
\end{equation}
In this limit, the fraction pumped of each $M$ is proportional to $\Omega_{12}^2$ and thus proportional to the intensity and the square of the transition matrix element.  

For non-saturated probe transitions, $\Omega_{23} << \gamma_1 \gamma_2$, the absorption coefficient of the probe is given by
\begin{equation}
\alpha_{23}(\Delta \omega_{23}) = \frac{\omega_{23}}{\epsilon_0 c \hbar} \frac{\gamma_2}{\gamma_2^2 + \Delta \omega_{23}^2} \cdot \sum_{M = -J_2}^{J_2}  \mu_{23}(M)^2  \Delta \rho_{22}(M) 
\end{equation}
Integrating over the probe detuning, we get an integrated absorption coefficient we get for the unsaturated case:
\begin{equation}
I_G = \int \alpha_{23} d\omega_{23} =  \frac{\pi^2 \omega_{23}}{ (J_2+1) (\epsilon_0 c)^{2} \hbar^3 \gamma_1}  ( \rho_{11}^e - \rho_{22}^e )  g_{\rm D}(\omega - \omega_{12}) 
 I_p  \cdot \sum_{M = -J_2}^{J_2}  \mu_{23}(M)^2    \mu_{12}(M)^2 
\end{equation}
Using the axes assignments given above, for symmetric top transitions $J_1, K_1, M \rightarrow J_2, K_2, M \rightarrow J_3, K_3, M'$ we have
\begin{eqnarray}
 \sum_{M = -J_2}^{J_2}  \mu_{23}(M)^2    \mu_{12}(M)^2 & &=  \mu_{12}^2 \, \mu_{23}^2 \, \phi_J(J_1,J_2)^2 \phi_g(J_1,K_1,J_2, K_2)^2  
 \phi_J(J_2,J_3)^2 \phi_{g'}(J_2, K_2, J_3, K_3)^2  \times \nonumber \\
&&  \sum_{M = -J_2}^{J_2} \phi_Z(J_1,M,J_2, M)^2 \sum_{M'} \phi_G(J_2,M,J_3,M')^2 \label{mu12mu23}
\end{eqnarray}
In this equation, $\mu_{12}$ and $\mu_{23}$ are the transition dipole moment matrix elements in the molecular frame for the pump and probe transitions, and $G = Z$ or $X$ depending on whether the probe is polarized parallel to the pump, giving a signal $I_{\parallel}$ or perpendicular to the pump, giving a signal $I_{\perp}$.  When $G = Z$, the selection rule gives $M' = M$, and when $G = X$, the selection rule $M' = M \pm 1$.

For asymmetric top molecules, we can expand the rotational wavefunction for each state with quantum numbers $i, J_i, \tau_i, M$ as $\phi(i, J_i, \tau_i, M)  = \sum_K A(i, J_i, \tau_i, K) \phi_{J_i,K,M}$ where $\phi_{J_1,K,M}$ are symmetric top wavefunctions.  In Eq.~\ref{mu12mu23}, we replace the terms $\phi_g(J_i, K_i, J_j, K_j)^2$ by $ | \sum_{K_i, K_j}  A(i, J_i, \tau_i, K_i, )  A(j, J_j, \tau_j,  K_j )|^2 \phi_g(J_i, K_i, J_j, K_j) |^2$,  

The DR signal strength for an arbitrary angle, $\theta$, between pump and probe polarizations can be written as $I(\theta) = I_{\parallel} \cos^2 \theta + I_{\perp} \sin^2(\theta)$.  For $\theta_m = \cos^{-1}(1/\sqrt{3})$ (known as the magic angle) $I(\theta_m) = (I_{\parallel} + 2 I_{\perp})/3$.  The sum $\phi_Z(J,M,J',M)^2 + 2 \phi_X(J,M,J',M+1)^2 + 2\phi_X(J,M,J',M-1)^2$ is $M$ independent and so, at the magic angle of relative polarization, the probe absorption strength is proportional to the total population in the intermediate energy level $J_2, K_2$.

The ratio of $I_{\parallel}$ and $I_{\perp}$ depends only on the sum over $M$ values as all other factors are independent of pump or probe
polarization, thus the polarization ratio in the unsaturated pump case is
\begin{equation}
R^{\rm us}_{\rm lin} = \frac{I_{\parallel}}{I_{\perp}} = \frac{ \sum_{M = -J_2}^{J_2} \phi_Z(J_1,M,J_2.M)^2  \phi_Z(J_2,M,J_3,M)^2     }{ 
 \sum_{M = -J_2}^{J_2}   \phi_Z(J_1,M,J_2.M)^2  ( \phi_X(J_2,M,J_3,M-1)^2 + \phi_X(J_2,M,J_3,M+1)^2)   }  \label{eq:usLinIntRatio}
\end{equation}
Given the $\Delta J = 0, \pm 1$ selection rule for both pump and probe transition, there are 9 possible cases and the ratio given in Eq.~\ref{eq:usLinIntRatio} can be evaluated using the expressions for $\phi_G$ given in table~\ref{DirCosineElements}.
The resulting analytical expressions (with sums over $M$ evaluated using Mathematica) are presented in Table~\ref{LinPol},  both in symbolic form and numerical values for $J_2 = 0-10$.   It is traditional to label molecular transitions with  $R, Q, P$ for transitions when  $J$ for the upper state minus $J$ for the lower states $= +1, 0, -1$\, respectively.
These labels change for the three different DR schemes: (ladder-type with $E_1 < E_2 < E_4$, V-type with $E_2 < E_1, E_2$, and $\Lambda$-type with $E_2 > E_1, E_3$.  Missing entries in the table correspond to dipole forbidden transitions which require at least one of the two $J$ values for each transition to be greater than $0$.

\begin{table}[htp]
\caption{Table of nonzero direction cosine matrix elements taken from Townes \& Schawlow.  The symmetric top matrix element of $< J,K,M | \mu_g E_h | J', K', M' > =  \phi_(J,J') \phi_g(J,K,J',K')^{*} \phi_h(J,M,J',M')$
where $g$ is the direction of the transition moment in the molecular axis system and $h$ is the direction of the electric field in the laboratory fixed axis system. The root mean square values 
$\phi_{\rm rms}^2 = \sum_{M = -J}^{J}  \phi_Z(J,M,J',M)^2 / (2J+1) =  2 \sum_{M = -J}^{J}  \phi_X(J,M,J',M \pm 1)^2 / (2J+1)  =  2 \sum_{M = -J}^{J} | \phi_Y(J,M,J',M \pm 1) |^2 / (2J+1) $  }
\begin{center}
\begin{tabular}{| c | c |  c |  c |  }
\hline
                                         &  $ J' = J +1 $   &  $ J' = J  $   &  $ J' = J -1 $   \\  \hline
$ \phi(J,J') $                     &    $ \left[  4 (J+1) \sqrt{ (2J+1)(2J+3)} \right]^{-1} $   & $ \left[  4 J(J+1)  \right]^{-1}  $  & $ \left[  4 J \sqrt{ (2J-1)(2J+1  )} \right]^{-1}  $  \\
$\phi_Z(J,M,J',M)  $         &     $  2 \sqrt{(J+1)^2 - M^2}       $  &    $ 2 M $     &  $ 2 \sqrt{J^2 - M^2}       $  \\ 
$(\phi_X {\rm \,\, or \pm \phi_Y}) (J,M,J',M \pm 1)$ &  $ \mp \sqrt{ (J \pm M +1) (J  \pm M  + 2)  } $   &  $ \sqrt{ (J \mp M) (J  \pm M  + 1)  } $   &  $ \pm \sqrt{ (J \mp M +1) (J  \mp M  -1)  } $ \\
$ \phi_{\rm rms}(J,J')^2 $    & $ \frac{4}{3} (J+1)(2J+3)  $  &  $ \frac{4}{3} J(J+1)  $  &  $ \frac{4}{3} J(2J-1)  $ \\
\hline
\end{tabular}
\end{center}
\label{DirCosineElements}
\end{table}

\begin{table}[htp]
\caption{Ratio of double resonance signals for parallel over perpendicular relative linear polarization for unsaturated double resonance.  J is the rotational
total angular momentum quantum number for the state common to the two transitions }
\begin{center}
\begin{tabular}{|r|r| l| l| l| l| l| l| l| l| l|}
\hline
ladder-type         & R then R    & R then Q   & R then P    & Q then R    & Q then Q & Q then P  & P then R   & P then Q   & P then P \\ 
V-type                 & P then R    & P then Q   & P then P    & Q then R    & Q then Q  & Q then P  & R then R  & R then Q   & R then P \\ 
$\Lambda$-type &  R then P   & R then Q   & R then R   & Q then P     & Q then Q & Q then R  & P then P   & P then Q   & P then R \\ 
$J_1 =$              &  $J_2 - 1$  &  $J_2 - 1$ &  $J_2 - 1$  &  $J_2 $       & $J_2 $     & $J_2 $     & $J_2 + 1$ & $J_2 + 1$ & $J_2 + 1$ \\ 
$J_3 =$              &   $J_2 +1$ &  $J_2 $     &  $J_2 - 1$  &  $J_2 +1 $  & $J_2 $     & $J_2 -1 $ & $J_2 + 1$ & $J_2      $ & $J_2 - 1$ \\ 
\hline
       $ J_2 $              & $\frac{4}{3}$ & $\frac{2 (J_2-1)}{4J_2+1}$  & $ \frac{8 J_2^2 + 2}{(J_2-1)(6J_2+1)} $ & $ \frac{2J_2+4}{4J_2+3} $ &
$ \frac{6 J_2^2 + 6J_2 - 2}{ 2 J_2^2 + 2J_2 + 1   }  $  &$ \frac{ 2J_2-2  }{4J_2+1   } $ &
$\frac{8 J_2^2 + 16J_2 + 10  }{ (J_2+2)(6J_2+5)  } $ & $\frac{  2J_2+4}{4J_2+3 } $ & $\frac{4}{3}$ \\ 
\hline
0   &                 &                 &                 &                 &                 &                 &   1.   &                 &     \\
1   &   1.3333   &   0.            &   $\infty$   &   0.8571   &   2.   &   0.            &   1.0303   &   0.8571   &   1.3333    \\
2   &   1.3333   &   0.2222   &   2.6154   &   0.7273   &   2.6154   &   0.2222   &   1.0882   &   0.7273   &   1.3333    \\
3   &   1.3333   &   0.3077   &   1.9474   &   0.6667   &   2.8000   &   0.3077   &   1.1304   &   0.6667   &   1.3333    \\
4   &   1.3333   &   0.3529   &   1.7333   &   0.6316   &   2.8780   &   0.3529   &   1.1609   &   0.6316   &   1.3333    \\
5   &   1.3333   &   0.3810   &   1.6290   &   0.6087   &   2.9180   &   0.3810   &   1.1837   &   0.6087   &   1.3333    \\
6   &   1.3333   &   0.4000   &   1.5676   &   0.5926   &   2.9412   &   0.4000   &   1.2012   &   0.5926   &   1.3333    \\
7   &   1.3333   &   0.4138   &   1.5271   &   0.5806   &   2.9558   &   0.4138   &   1.2151   &   0.5806   &   1.3333    \\
8   &   1.3333   &   0.4242   &   1.4985   &   0.5714   &   2.9655   &   0.4242   &   1.2264   &   0.5714   &   1.3333    \\
9   &   1.3333   &   0.4324   &   1.4773   &   0.5641   &   2.9724   &   0.4324   &   1.2357   &   0.5641   &   1.3333    \\
10  &  1.3333   &   0.4390   &   1.4608   &   0.5581   &   2.9774   &   0.4390   &   1.2436   &   0.5581   &   1.3333    \\
\hline
\end{tabular}
\end{center}
\label{LinPol}
\end{table}

In the unsaturated limit, the polarization ratios are the same for homogeneously and inhomogeneously broadened cases.
We are not aware in the literature of explicit expressions for the predicted DR polarization intensity ratio, $I_{\parallel}/I_{\perp}$, for all cases. 
These expressions in Table~\ref{LinPol}  can be derived from those previously given
in Table~5.1 of the text \textit{Angular Momentum} by Richard Zare,\cite{ZareAM} which gives the degree of polarization for fluorescence, which is analogous to the $\Lambda$-type DR scheme.  
Zare gave expressions for the polarization anisotropy defined as $P = (  	I_{\parallel}  - I_{\perp})/(I_{\parallel}  + I_{\perp})$.   The expressions for $P$ were converted by using $I_{\parallel} / I_{\perp} = (P+1)/(P-1)$ and found to agree with those given in Table~\ref{LinPol}, after correcting for the fact that Zare used what we have written as $J_1$ for his expressions and
we have used $J_2$,  as that allows all three DR schemes to be combined.  
  If the sample is optically thick for the pump, the polarization ratio, $R_{\rm lin}$, will not change.  If the probe is optically thick, then $R_{\rm lin}$ will give the ratio of the change in the probe field absorbance induced by the pump field.

It is also possible to do the polarization measurement using circular polarization for the pump and probe fields.\cite{Schwendeman99}  In this case, we assign the $Z$ axis to the direction of propagation of the pump field and that (or it's inverse) of the probe field.     The pump and probe fields will be assumed to be circularly polarized with positive or negative helicity and the propagation
direction, $\vec{k}$.    The nonzero direction cosine matrix elements are $\phi_{\pm}(J,M,J',M\pm 1 ) = (\phi_X (J,M,J',M\pm 1 )  \pm i \phi_Y(J,M,J',M\pm 1)) /\sqrt{2} = \sqrt{2} \phi_X (J,M,J',M\pm 1 )$.
The positive sign is used for absorption from a wave of positive helicity or stimulated emission from a wave of negative helicity.  If the wave propagates with $\vec{k}$ antiparallel to the $Z$ axis, the signs are switched.
We compare the predicted DR signal strength for the cases where the pump and probe co-rotate together or counter-rotate, with signal strengths $I_{\rm same}$ and $I_{\rm opposite}$ respectively.  Note that if the pump and probe waves have parallel $\vec{k}$, the waves have the same or opposite helicities for co-rotating and counter-rotating respectively.   However, if they have antiparallel $\vec{k}$, then co- and counter rotation corresponds to the helicities being opposite or the same respectively.  The helicity of a wave is reversed upon reflection at normal incidence from a mirror, so one can use a double pass or even a linear enhancement cavity to increase the DR signal strength assuming the mirrors have negligible birefringence and dichroism.  Using the direction cosine matrix elements, one can evaluate the polarization ratios for ladder- and V-type DR as
\begin{equation}
R^{\rm us}_{\rm cir} = \frac{I_{\rm same}}{I_{\rm opposite} } = \frac{  \sum_{M = -J_2}^{J_2}  \phi_X(J_1,M-1,J_2.M)^2  \phi_X(J_2,M,J_3,M+1)^2    }{ 
  \sum_{M = -J_2}^{J_2}   \phi_X(J_1,M-1,J_2,M)^2   \phi_X(J_2,M,J_3,M-1)^2   }  \label{eq:usCirIntRatio}
\end{equation}
As the change in the $M$ quantum number for stimulate emission is opposite that for absorption for fixed helicity, $R^{\rm us}_{\rm cir}$ for $\Lambda$-type DR will be the inverse of that
given in Eq.~\ref{eq:usCirIntRatio}.  
Table~\ref{CirPol}  gives the predicted non-saturated DR circular polarization ratios.    It is evident that the polarization effects are generally larger for the comparison of circular vs linear polarization DR signals.  In particular, for linear polarization, the predicted ratios for probe transitions with $\Delta J = \pm 1$ approach each other as $J$
 grows, making discrimination difficult.   For circular polarization, the two $\Delta J = \pm 1$ probe transitions are most easily distinguished.   
 
 Zare's text gives the degree of circular polarization,  $C(J_1) = (I_{\rm same} - I_{\rm opposite})/I_{\rm same} + I_{\rm opposite})$  for fluorescence in Table~5.2.  Correcting for the inversion for emission vs absorption ( equivalently, changing the sign of $C(J_1)$)~
 his expressions can be converted to those given in Table~\ref{CirPol} and are found to agree with the exception of the case he labels  $(R\uparrow \, Q\downarrow)$.  
 However, recalculating that $C(J_1)$ value using his eq.~5.124 shows that there was a sign error in his printed table for that entry.
  With this correction, the two expressions are in agreement in that case as well.  

\begin{table}[htp]
\caption{Double resonance signal polarization ratios using circularly polarized radiation for unsaturated transitions.  For ladder- and V-type double resonance, what is tabulated is same helicity / opposite helicity.
For $\Lambda$-type DR, what is tabulated is opposite helicity / same helicity.  J is the rotational
total angular momentum quantum number for the state common to the two transitions  }
\begin{center}
\begin{tabular}{|r|r| l| l| l| l| l| l| l| l|}
\hline
ladder-type           & R then R  & R then Q  & R then P   & Q then R & Q then Q & Q then P & P then R & P then Q & P then P \\ 
V-type                   & P then R  & P then Q  & P then P   & Q then R & Q then Q & Q then P  & R then R & R then Q & R then P \\ 
$\Lambda$-type   &  R then P & R then Q  & R then R  & Q then P & Q then Q  & Q then R & P then P  & P then Q & P then R \\ 
$J_1 =$ &  $J_2 - 1$ &  $J_2 - 1$ &  $J_2 - 1$ &  $J_2 $      & $J_2 $ & $J_2 $     & $J_2 + 1$ & $J_2 + 1$ & $J_2 + 1$ \\ 
$J_3 =$ &   $J_2 +1$ &  $J_2 $     &  $J_2 - 1$ &  $J_2 +1 $ & $J_2 $ & $J_2 -1 $ & $J_2 + 1$ & $J_2      $ & $J_2 - 1$ \\ 
\hline
$J_2$ & 6   &  $\frac{3J_2-3}{3J_2+2}$ &
$\frac{(J_2-1)(2J_2-3)}{ 12 J_2^2 - 2  }$ & $\frac{3J_2+6  }{ 3J_2 + 1} $ & $\frac{ (2J_2+3)(2J_2-1) }{4 J_2^2  +4J_2 + 2 }$ &
$\frac{ 3J_2-3 }{3J_2+2}$ &  $\frac{(J_2+2)(2J_2+5)}{12J_2^2 +24 J_2 + 10}$   &   $\frac{3J_2+6}{3J_2+1}$   &   6. \\ \hline
0   &       &            &            &            &            &            &   1.            &            &      \\
1   &   6   &   0.   &   0.    &   2.2500  &   0.5   &   0.  &   0.4565   &   2.2500   &  6  \\
2   &   6   &   0.3750   &   0.0217   &   1.7143   &   0.8077   &   0.3750   &   0.3396   &   1.7143   &   6  \\
3   &   6   &   0.5455   &   0.0566   &   1.5000   &   0.9000   &   0.5455   &   0.2895   &   1.5000   &   6  \\
4   &   6   &   0.6429   &   0.0789   &   1.3846   &   0.9390   &   0.6429   &   0.2617   &   1.3846   &   6  \\
5   &   6   &   0.7059   &   0.0940   &   1.3125   &   0.9590   &   0.7059   &   0.2442   &   1.3125   &   6  \\
6   &   6   &   0.7500   &   0.1047   &   1.2632   &   0.9706   &   0.7500   &   0.2321   &   1.2632   &   6  \\
7   &   6   &   0.7826   &   0.1126   &   1.2273   &   0.9779   &   0.7826   &   0.2232   &   1.2273   &   6  \\
8   &   6   &   0.8077   &   0.1188   &   1.2000   &   0.9828   &   0.8077   &   0.2165   &   1.2000   &   6  \\
9   &   6   &   0.8276   &   0.1237   &   1.1786   &   0.9862   &   0.8276   &   0.2112   &   1.1786   &   6  \\
10 &   6   &   0.8438   &   0.1277   &   1.1613   &   0.9887   &   0.8438   &   0.2069   &   1.1613   &   6  \\
\hline 
\end{tabular}
\end{center}
\label{CirPol}
\end{table}

\subsection{Saturated Inhomogeneously broadened Pump Transitions}
 
 In steady-state, saturation of the pump and/or probe transitions will generally reduce the polarization effects as it will reduce
the degree of alignment produced by the pump beam and/or reduce the impact of alignment on
the probe absorption.   We will only consider the case with saturation of the pump transition.
We start by considering the case where the inhomogeneous Doppler width is the dominant broadening.
This is the common situation in continuous wave excitation experiments due to the fact it is difficult to realize a Rabi
excitation frequency greater than the Doppler width of the transition, at least for thermal samples at
near ambient temperatures or above.   
 In the limit of a strongly saturating pump wave, $\Omega_{12}^2(M) >> \gamma_1 \gamma_2$, for all nonzero values of $\mu_{12}(M)$.  We again will assume that the inhomogeneous Doppler width dominates over $\Omega_{12}, \gamma_1$ and $\gamma_2$.  Returning to Eq.~\ref{eq:drho2} and taking this limit, we have for the total change in population in state 2:
\begin{equation}
\Delta \rho_{22}(M) \rightarrow  \frac{\pi}{2 (2 J_2 +1)}  ( \rho_{11}^e - \rho_{22}^e )  g_{\rm D}(\omega_a - \omega_{12})  | \Omega_{12}(M) | \sqrt{\gamma_2 / \gamma_1) }
\end{equation}
and the population transferred to level $2$ is proportional to the $ | \Omega_{12}(M) |$ and thus the square root of the pump intensity.
We thus get the following sums for DR transitions $J_1, K_1 \rightarrow J_2, K_2, \rightarrow J_3, K_3 $ of symmetric tops
\begin{eqnarray}
 \sum_{M = -J_2}^{J_2}  \mu_{23}(M)^2   | \mu_{12}(M) | & &=  |\mu_{12}| \, \mu_{23}^2 \, \phi_J(J_1,J_2) |\phi_g(J_1,K_1,J_2, K_2)| 
 \phi_J(J_2,J_3)^2 \phi_{g'}(J_2,K_2,J_3, K_3)^2  \times \nonumber \\
&&  \sum_{M = -J_2}^{J_2} | \phi_Z(J_1, M, J_2, M) | \sum_{M'} \phi_G(J_2,M,J_3,M')^2
\end{eqnarray}
As above, the asymmetric top case can be found by replacing the terms $|\phi_g(J_i,K_i, J_j, K_j ) |$
by $ | \sum_{K_i, K_j}  A(i, J_i, \tau_i, K_i, )  A(j, J_j, \tau_j,  K_j )|^2 \phi_g(J_i, K_i, J_j, K_j) |$.
These results leads to the expressions, in the limit of strong pump saturation,
\begin{equation}
I^{\rm sat}_G =   \frac{\pi^2 \omega_{23}}{2 (J_2+1) (\epsilon_0 c)^{3/2} \hbar^2}  ( \rho_{11}^e - \rho_{22}^e )  g_{\rm D}(\omega - \omega_{12}) 
\sqrt{\gamma_2 / \gamma_1) } \sqrt{2 I_p}  \cdot \sum_{M = -J_2}^{J_2}  \mu_{23}(M)^2   | \mu_{12}(M) | 
\end{equation}
and
\begin{equation}
R^{\rm sat}_{\rm lin} = \frac{I_{\parallel}}{I_{\perp}} = \frac{ \sum_{M = -J_2}^{J_2} \phi_Z(J_1,M,J_2.M)  \phi_Z(J_2,M,J_3,M)^2     }{ 
 \sum_{M = -J_2}^{J_2}   \phi_Z(J_1,M,J_2.M) ( \phi_X(J_2,M,J_3,M-1)^2 + \phi_X(J_2,M,J_3,M+1)^2)   }  \label{eq:LinIntRatio}
\end{equation}
\begin{equation}
R^{\rm sat}_{\rm cir} = \frac{I_{\rm same}}{I_{\rm opposite} } = \frac{  \sum_{M = -J_2}^{J_2}  \phi_X(J_1,M-1,J_2.M)  \phi_X(J_2,M,J_3,M+1)^2    }{ 
  \sum_{M = -J_2}^{J_2}   \phi_X(J_1,M-1,J_2,M)   \phi_X(J_2,M,J_3,M-1)^2   }  \label{eq:CirIntRatio}
\end{equation}
 
Due the fact that most of the factors $|\phi_Z(J_1,M,J_2,M)|$ and 
$|\phi_X(J_1,M, J_2, M + 1)|$ are square roots of polynomials in
$J_2$ and $M$, the sums over $M$ values do not lead to compact
expressions in the saturated case.   The exceptions are for Q pump ratios for linear polarization.
 However, it is straightforward to numerically calculate
the relevant sums for any particular values of $J_1, J_2$, and $J_3$.  
This has been done and the results for linear and circular polarization
are presented in tables~\ref{SatLinPol} and \ref{SatCirPol}.  In the comparison 
of the tables for unsaturated and saturated pump conditions, it is evident
that, while pump saturation reduces the polarization effects, this reduction is
modest and polarization can be used to unambiguously assign transitions.  
The polarization ratios are independent of the optical depth of the pump
transition as long as the steady-state excitation remains strongly
saturated through the sample.

 \begin{table}[h]
\caption{Ratio of double resonance signals for parallel over perpendicular relative linear polarization for strongly saturated inhomogeneously broadened pump transitions
but unsaturated probe transitions.  J is the rotational
total angular momentum quantum number for the state common to the two transitions.  The first transition listed is the pump' and the 2nd the probe. }
\begin{center}
\begin{tabular}{|r|l| l| l| l| l| l| l| l| l|}
\hline
ladder-type         & R then R    & R then Q   & R then P  & Q then R    & Q then Q & Q then P & P then R    & P then Q   & P then P \\ 
V-type                 & P then R    & P then Q   & P then P   & Q then R    & Q then Q & Q then P & R then R   & R then Q   & R then P \\ 
$\Lambda$-type &  R then P   & R then Q   & R then R  & Q then P    & Q then Q & Q then R & P then P    & P then Q   & P then R \\ 
$J_1 =$              &  $J_2 - 1$  &  $J_2 - 1$ &  $J_2 - 1$ &  $J_2 $      & $J_2 $     & $J_2 $     & $J_2 + 1$ & $J_2 + 1$ & $J_2 + 1$ \\ 
$J_3 =$              &   $J_2 +1$ &  $J_2 $     &  $J_2 - 1$ &  $J_2 +1 $ & $J_2 $     & $J_2 -1 $ & $J_2 + 1$ & $J_2      $ & $J_2 - 1$ \\ 
\hline
$J_2$  &    &    &    &  $\frac{2J_2+4}{ 3J_2+4  }  $ &  2.    & $ \frac{ 2J_2-2 }{ 3J_2 -1   }  $  &   &   &    \\  \hline
0   &           &                 &                 &            &            &                 &     1.      &                 &               \\  
1   &   1.3333   &   0.           &   $\infty$  &   0.8571   &   2.   &   0.            &   1.0148   &   0.9282   &   1.1547  \\
2   &   1.3244   &   0.2363   &   2.5559   &   0.8000   &   2.   &   0.4          &   1.0449   &   0.8527   &   1.1634  \\
3   &   1.3038   &   0.3514   &   1.8494   &   0.7692   &   2.   &   0.5          &   1.0682   &   0.8125   &   1.1690  \\
4   &   1.2868   &   0.4194   &   1.6207   &   0.7500   &   2.   &   0.5455   &   1.0856   &   0.7873   &   1.1703  \\
5   &   1.2739   &   0.4640   &   1.5089   &   0.7368   &   2.   &   0.5714   &   1.0990   &   0.7698   &   1.1759  \\
6   &   1.2640   &   0.4954   &   1.4431   &   0.7273   &   2.   &   0.5882   &   1.1096   &   0.7570   &   1.1783  \\
7   &   1.2562   &   0.5187   &   1.3998   &   0.7200   &   2.   &   0.6000   &   1.1182   &   0.7471   &   1.1802  \\
8   &   1.2500   &   0.5365   &   1.3694   &   0.7143   &   2.   &   0.6087   &   1.1253   &   0.7392   &   1.1817  \\
9   &   1.2450   &   0.5507   &   1.3468   &   0.7097   &   2.   &   0.6154   &   1.1312   &   0.7328   &   1.1830  \\
10 &   1.2408   &   0.5621   &   1.3294   &   0.7059   &   2.   &   0.6207   &   1.1362   &   0.7275   &   1.1842  \\

\hline
\end{tabular}
\end{center}
\label{SatLinPol}
\end{table}

\begin{table}[h]
\caption{Double resonance signal polarization ratios using circularly polarized radiation
for strongly saturated inhomogeneously broadened pump transitions
but unsaturated probe transitions.  For ladder- and V-type Double resonance, what is tabulated is the same helicity / opposite helicity.
For $\Lambda$-type DR, what is tabulated is opposite helicity / same helicity.  J is the rotational
total angular momentum quantum number for the state common to the two transitions.  The first transition listed is the pump; the 2nd the probe. }
\begin{center}
\begin{tabular}{|r|l| l| l| l| l| l| l| l| l|}
\hline
ladder-type         & R then R    & R then Q   & R then P  & Q then R     & Q then Q & Q then P & P then R   & P then Q   & P then P \\ 
V-type                 & P then R    & P then Q   & P then P   & Q then R   & Q then Q  & Q then P  & R then R  & R then Q   & R then P \\ 
$\Lambda$-type &  R then P   & R then Q   & R then R  & Q then P    & Q then Q  & Q then R  & P then P   & P then Q   & P then R \\ 
$J_1 =$              &  $J_2 - 1$  &  $J_2 - 1$ &  $J_2 - 1$ &  $J_2 $      & $J_2 $      & $J_2 $     & $J_2 + 1$ & $J_2 + 1$ & $J_2 + 1$ \\ 
$J_3 =$              &   $J_2 +1$ &  $J_2 $     &  $J_2 - 1$ &  $J_2 +1 $ & $J_2 $      & $J_2 -1 $ & $J_2 + 1$ & $J_2      $ & $J_2 - 1$ \\ 
\hline
$J_2 = 0 $  &           &                 &                 &                 &                 &                 &   1.   &                 &              \\
1   &   6.           &    0.          &   0.           &   2.25   &   0.5         &   0.           &   0.6531   &   1.5306   &   2.4495 \\
2   &   4.4016   &   0.4936   &   0.0479   &   1.7071   &   0.8040   &   0.3876   &   0.5459   &   1.3439   &   2.5777 \\
3   &   3.8908   &   0.6628   &   0.1234   &   1.4914   &   0.8967   &   0.5582   &   0.4930   &   1.2549   &   2.6562 \\
4   &   3.6454   &   0.7477   &   0.1701   &   1.3761   &   0.9365   &   0.6542   &   0.4614   &   1.2028   &   2.7095 \\
5   &   3.5030   &   0.7985   &   0.2005   &   1.3046   &   0.9571   &   0.7159   &   0.4403   &   1.1684   &   2.7482 \\
6   &   3.4106   &   0.8323   &   0.2216   &   1.2559   &   0.9691   &   0.7589   &   0.4252   &   1.1440   &   2.7776 \\
7   &   3.3461   &   0.8564   &   0.2371   &   1.2206   &   0.9767   &   0.7906   &   0.4139   &   1.1258   &   2.8008 \\
8   &   3.2986   &   0.8745   &   0.2489   &   1.1938   &   0.9818   &   0.8149   &   0.4051   &   1.1117   &   2.8195 \\
9   &   3.2623   &   0.8885   &   0.2581   &   1.1729   &   0.9854   &   0.8341   &   0.3980   &   1.1004   &   2.8350 \\
10  &   3.2337   &   0.8997   &   0.2656  &   1.1560   &   0.9880   &   0.8497   &   0.3922   &   1.0912   &   2.8480 \\
\hline
\end{tabular}
\end{center}
\label{SatCirPol}
\end{table}

\subsection{ Comparison with Experimental Data}
The author and collaborators have performed IR-IR double resonance experiments in the spectral range of the CH$_4$ ground $\rightarrow \nu_3 \rightarrow 3\nu_3$.
A CW 3.3\,$\mu$m Optical Parametric Oscillator was used for the pump, and a 1.65\,$\mu$m centered frequency comb as the probe.
The pump was strongly saturated with the pump Rabi frequency about an order of magnitude larger than the collisional dephasing rate.  A preliminary
experiment using a single pass, liquid N$_2$ cooled cell has been published.\cite{Foltynowicz21a, Foltynowicz21b}  The method has greatly improved sensitivity
by using a finesse optical cavity for the probe radiation.\cite{deoliveira2023subdoppler}  Table~\ref{Exp_values_table} presents a set of
measured double resonance transitions from this latter work, each measured with both parallel and perpendicular relative polarization of the pump and probe waves.
The probe $\Delta J$ values were assigned based on combination differences and also by comparison with highly accurate theoretical predictions.\cite{Rey18}
The observed and calculated polarization ratios have significant quantitative differences,  with observed
values systematically closer to unity.  Such a bias towards unity can be expected from errors in the relative polarization state of the pump and probe lasers; as that
experimental setup has been replaced, we will not now speculate on the specific reason(s) for the deviation of the experimental polarization ratios from the predicated values.
Despite this, the saturated wave predictions are generally closer to the observed values.   
This demonstrates that even with strong saturation of the pump transitions, which optimizes detection sensitivity, polarization ratios can still be used to unambiguously assign the $\Delta J$ for the observed DR transitions.

\begin{table}[htp]
\caption{Comparison of observed and predicted polarization intensity ratios for Methane ground state $\rightarrow \,\, \nu_3 \,\, \rightarrow 3\nu_3$\,\, double resonance 
intensity ratios.  Experimental values taken from de Oliveria {\it et al.}\cite{deoliveira2023subdoppler}  The probe intensity is the integrated probe absorption of the sub-Doppler feature in units of 10$^{-9}$\, cm$^{-2}$.}
\begin{center}
\begin{tabular}{|l|l|l|l|l|l|l|l|l}
\hline  
Pump & Probe & Wavenumber & Final Term Value & Probe     & Polarization & Saturation & \text{Unsaturated} \\
trans  & trans   & cm$^{-1}$      &  cm$^{-1}$           &  Intensity& Ratio           & Prediction &  \text{Prediction}   \\
\hline
P(2F2) & R(1)   & 5948.267590(3)    & 8978.704010(3)  & 2.26(5)   & 0.91(4)      &    1.0148  &  1.0303      \\
P(2F2) & R(1)   & 5964.06227(2)      & 8994.49869(2)   & 0.081(3)  & 1.01(7)      &    1.0148  &    1.0303    \\
P(2F2) & R(1)   & 5979.042972(3)    & 9009.479392(3) & 0.56(1)    & 0.98(4)      &    1.0148  &   1.0303     \\
Q(2F2) & Q(2F1) & 5928.61142(2)   & 8978.70401(2)    & 0.56(3)    & 1.50(10)    &    2.0000  &   2.6154     \\
Q(2F2) & Q(2F1) & 5944.40608(2)   & 8994.49868(2)    & 0.103(3)  & 1.55(6)      &    2.0000  &   2.6154       \\
Q(2F2) & R(2F1) & 5958.673574(6) & 9008.766169(6)  & 2.38(6)    & 0.83(5)      &   0.8000  &   0.7273     \\
Q(2F2) & Q(2F1) & 5959.386797(5) & 9009.479392(5)  & 0.89(2)    & 1.84(7)      &   2.0000  &   2.6154       \\
R(2F2) & R(3F1) & 5913.18732(2)   & 8992.78303(2)    & 0.062(4)   & 1.1(1)       &  1.3038   &   1.3333     \\
R(2F2) & R(3F1) & 5918.14141(1)   & 8997.73712(1)   & 0.096(4)    & 1.05(7)     &   1.3038  &   1.3333      \\
R(2F2) & R(3F1) & 5923.94848(1)   & 9003.54418(1)   & 0.44(2)      & 1.19(6)     &   1.3038  &    1.3333   \\
R(2F2) & R(3F1) & 5924.26536(2)   & 9003.86107(2)   & 0.175(5)    & 1.10(8)     &   1.3038  &    1.3333    \\
R(2F2) & Q(3F1) & 5929.170466(3) & 9008.766172(3) & 4.7(1)        & 0.44(7)     &   0.3514  &    0.3077    \\
R(2F2) & P(3F1) & 5929.883687(4) & 9009.479393(4) & 0.85(2)      & 1.61(4)      &   1.8494  &     1.7333    \\
R(2F2) & R(3F1) & 5932.279186(9) & 9011.874892(9) & 0.100(3)    & 1.21(9)      & 1.3038    &   1.3333    \\
R(2F2) & R(3F1) & 5935.245195(3) & 9014.840901(3) & 0.74(2)      & 1.36(5)      &   1.3038  &    1.3333    \\
\hline
\end{tabular}
\end{center}
\label{Exp_values_table}
\end{table}

\subsection{ Double Resonance Polarization Ratios for Saturated and Homogeneously Broadened pump transitions}

For the sake of completeness, we now consider the saturation of saturated and homogeneously dominated broadened pump transitions.
As mentioned above, for unsaturated transitions, the same polarization ratios apply to the homogeneous and inhomogeneously
broadened cases.
In the homogeneously broadened case, the steady-state $\Delta \rho_{22}(M)$ is proportional to  $x(M) / 2(1+x(M))$ with 
$x(M) = S\, \phi_Z(J_1,M,J_2,M)^2 / \phi_{\rm rms}(J_1,J_2)^2$ in the linear pump polarization case and 
$x = 2 \, S \, \phi_X(J_1,M,J_2,M)^2/ \phi_{\rm rms}(J_1,J_2)^2$  for circular pump polarization.
 $S$ is the saturation parameter which equals the ratio of the pump rate neglecting saturation divided by the population
relaxation rate and
$\phi_{\rm rms}(J_1,J_2)^2 = \sum_{M = -J_2}^{J_2} \phi_Z(J_1,M,J_2,M)^2 / ( 2 J_2 + 1) 
= 2 \sum_{M = -J_2}^{J_2}  \phi_X(J_1,M-1,J_2,M)^2  / ( 2 J_2 + 1) $.  Expressions for $\phi_{\rm rms}$ are
given in Table~\ref{DirCosineElements}.

Even in the limit of highly saturating
pump intensity, the polarization ratios do not go to unity except for a $J_1 = J_2 + 1$ pump transition as the selection rules for 
the $J_1 = J_2$ or $J_2-1$ cases prevent pumping all $M$ values of state $2$.  Below, in tables~\ref{InfiniteLinPol}
and \ref{InfiniteCirPol} are listed the $S \rightarrow \infty$ polarization ratios values for different cases.    Note
that the polarizations ratios in the limit of infinite saturation $\rightarrow 1$ as $ J \rightarrow \infty$ in all cases,
as in that limit, the non-pumped $M$ values are a negligible fraction of the total. Figures [1-3] plot the linear polarization
ratios as a function of $S$ for the three probe transitions when a homogeneously broadened $R(5), P(5)$, and $Q(5)$ pump is used.
Figures [4-6] plot the circular polarization ratios for the same pump transitions. Table~\ref{LinHomHalfSat} and \ref{CirHomHalfSat}
report the values of the saturation parameter, $S$, that results in a polarization ratio halfway between the unsaturated and $S \rightarrow \infty$
values for the linear and circular polarization DR experiments respectively,   Note that the $J_2 = 1, J_1 = J_3 = 0$ entry is empty for the 
linear polarization case as the polarization ratio, in that case, is $\infty$ regardless of $S$

\begin{table}[htp]
\caption{Polarization DR signal ratios for parallel over perpendicularly polarized waves for homogeneously broadened pump transition
in the limit of saturation parameter $\rightarrow \infty$.  $J$ is the total angular momentum quantum number of
the state common in both transitions. }
\begin{center}
\begin{tabular}{|c|c| c| c| cl }
\hline
        & $J_2 - J_1 =  1$  & $J_2 - J_1 = 0$   &  $J_2 - J_1 = -1 $  \\ 
\hline
$J_3 - J_2 = 1$ &    $\frac{ (2J_2+1)(J_2+3)  }{ 2 J_2^2+4J_2+3  } $    &   $\frac{ 8J_2+10  }{ 8J_2+13  } $ &   1   \\
$J_3 - J_2 = 0$   &   $\frac{ J_2-1 }{ J_2+2}$   &   $\frac{ 4J_2+2  }{ 4J_2-1  } $  &   1 \\
$J_3 - J_2 = -1 $ &    $\frac{J_2+1}{2J_2-2}$     &   $\frac{ 2(J_2-1)(4J_2+1)  }{ 8J_2^2 -3J_2 +1  } $ &   1  \\
\hline
\end{tabular}
\end{center}
\label{InfiniteLinPol}
\end{table}

\begin{table}[htp]
\caption{Polarization DR signal ratios for V- and ladder-type for co- over counter-rotating waves for Homogeneously broadened pump transition
in the limit of saturation parameter $\rightarrow \infty$.  $J$ is the total angular momentum quantum number of
the state common in both transitions.  For $\Lambda$-type DR, the ratio should be inverted.}
\begin{center}
\begin{tabular}{|c|c| c| c| cl }
\hline
         & $J_2 - J_1 = 1 $  & $J_2 - J_1 = 0 $ & $J_2 - J_1 = -1 $ \\  
\hline
$J_3 - J_2 = 1$  &     $\frac{2 J_2^2 + 7J_2 + 9}{(J_2(2J_2+1) }$    &  $\frac{4 J_2^2 + 12J_2 +  11}{(2J_2+1)(2J_2+1) }$    &   1 \\
$J_3 - J_2 = 0$   &    $\frac{(J_2-1)(J_2+3)}{J_2(J_2+2)}$                &  $\frac{(J_2+2)(2J_2-1)}{(J_2+1)(2J_2+1)}$                 &   1  \\
$J_3 - J_2 = -1$  &    $\frac{ (J_2-1)(2J_2-3) }{J_2(2J_2+1)}$         &  $\frac{ (2J_2-2) }{(2J_2+1)}$                                &   1  \\
\hline
\end{tabular}
\end{center}
\label{InfiniteCirPol}
\end{table}

\begin{table}[htp]
\caption{ Pump saturation parameter, $S$, required for the linear polarization ratio to be halfway between unsaturated and $S \rightarrow \infty$ limit }
\begin{center}
\begin{tabular}{|r|r| l| l| l| l| l| l| l| l|}
\hline 
ladder-type         & R then R    & R then Q   & R then P  & Q then R     & Q then Q & Q then P & P then R   & P then Q   & P then P \\ 
V-type                 & P then R    & P then Q   & P then P   & Q then R   & Q then Q  & Q then P  & R then R  & R then Q   & R then P \\ 
$\Lambda$-type &  R then P   & R then Q   & R then R  & Q then P    & Q then Q  & Q then R  & P then P   & P then Q   & P then R \\ 
$J_1 =$              &  $J_2 - 1$  &  $J_2 - 1$ &  $J_2 - 1$ &  $J_2 $      & $J_2 $      & $J_2 $     & $J_2 + 1$ & $J_2 + 1$ & $J_2 + 1$ \\ 
$J_3 =$              &   $J_2 +1$ &  $J_2 $     &  $J_2 - 1$ &  $J_2 +1 $ & $J_2 $      & $J_2 -1 $ & $J_2 + 1$ & $J_2      $ & $J_2 - 1$ \\ 
\hline
$J_2 = 1$	&2.0000	&2.0000	&		&1.9541	&2.0000	&2.0000	&0.5502	&0.5835	&1.2506  \\
2		&0.9213	&0.9377	&0.9030	&1.3278	&0.9288	&1.5002	&0.7924	&0.8975	&1.3969 \\
3		&1.1211	&1.1877	&1.0847	&1.7102	&1.0027	&1.9302	&0.9743	&1.1458	&1.5106 \\
4		&1.2688	&1.3904	&1.2250	&1.9762	&1.0431	&2.206	&1.1173	&1.3474	&1.6015 \\
5		&1.3840	&1.5583	&1.3374	&2.1697	&1.0703	&2.3947	&1.2334	&1.515	&1.676 \\
6		&1.4772	&1.7001	&1.4300	&2.3167	&1.0904	&2.5315	&1.3299	&1.6568	&1.7381 \\
7		&1.5544	&1.8216	&1.5076	&2.4321	&1.1062	&2.6353	&1.4116	&1.7787	&1.7908 \\
8		&1.6197	&1.9270	&1.5739	&2.5253	&1.1189	&2.7167	&1.4817	&1.8847	&1.8361 \\
9		&1.6758	&2.0195	&1.6312	&2.6022	&1.1294	&2.7824	&1.5427	&1.9779	&1.8755 \\
10		&1.7245	&2.1014	&1.6812	&2.6666	&1.1382	&2.8366	&1.5962	&2.0605	&1.9100 \\
\hline
\end{tabular}
\end{center}
\label{LinHomHalfSat}
\end{table}

\begin{table}[htp]
\caption{ Pump saturation parameter, $S$, required for the circular polarization ratio to be halfway between unsaturated and $S \rightarrow \infty$ limit }
\begin{center}
\begin{tabular}{|r|r| l| l| l| l| l| l| l| l|}
\hline 
ladder-type         & R then R    & R then Q   & R then P  & Q then R     & Q then Q & Q then P & P then R   & P then Q   & P then P \\ 
V-type                 & P then R    & P then Q   & P then P   & Q then R   & Q then Q  & Q then P  & R then R  & R then Q   & R then P \\ 
$\Lambda$-type &  R then P   & R then Q   & R then R  & Q then P    & Q then Q  & Q then R  & P then P   & P then Q   & P then R \\ 
$J_1 =$              &  $J_2 - 1$  &  $J_2 - 1$ &  $J_2 - 1$ &  $J_2 $      & $J_2 $      & $J_2 $     & $J_2 + 1$ & $J_2 + 1$ & $J_2 + 1$ \\ 
$J_3 =$              &   $J_2 +1$ &  $J_2 $     &  $J_2 - 1$ &  $J_2 +1 $ & $J_2 $      & $J_2 -1 $ & $J_2 + 1$ & $J_2      $ & $J_2 - 1$ \\ 
\hline
$J_2 = 1$	&2.0000	&2.0000	&2.0000	&1.9541	&2.0000	&2.0000	&0.5502	&0.5835	&1.2506  \\
2		&0.9213	&0.9377	&0.9030	&1.3278	&0.9288	&1.5002	&0.7924	&0.8975	&1.3969  \\
3		&1.1211	&1.1877	&1.0847	&1.7102	&1.0027	&1.9302	&0.9743	&1.1458	&1.5106   \\
4		&1.2688	&1.3904	&1.2250	&1.9762	&1.0431	&2.2060	&1.1173	&1.3474	&1.6015   \\
5		&1.3840	&1.5583	&1.3374	&2.1697	&1.0703	&2.3947	&1.2334	&1.5150	&1.6760   \\
6		&1.4772	&1.7001	&1.4300	&2.3167	&1.0904	&2.5315	&1.3299	&1.6568	&1.7381   \\
7		&1.5544	&1.8216	&1.5076	&2.4321	&1.1062	&2.6353	&1.4116	&1.7787	&1.7908   \\
8		&1.6197	&1.927	&1.5739	&2.5253	&1.1189	&2.7167	&1.4817	&1.8847	&1.8361   \\
9		&1.6758	&2.0195	&1.6312	&2.6022	&1.1294	&2.7824	&1.5427	&1.9779	&1.8755  \\
10		&1.7245	&2.1014	&1.6812	&2.6666	&1.1382	&2.8366	&1.5962	&2.0605	&1.9100 \\

\hline
\end{tabular}
\end{center}
\label{CirHomHalfSat}
\end{table}

\section{Summary and Conclusions}
This work has presented expressions that allow the prediction of changes in signal strength as a function of relative
pump and probe polarization applicable to a wide range of DR experiments performed with the pump population
transfer in the steady-state limit.  It is found that, even in the case of a strongly saturated pump field, most of
the polarization anisotropy remains in the case of an inhomogeneously broadened pump transition due to
the different power broadened widths of different $M$ projection states.  This allows
polarization ratios to be used to unambiguously assign the $\Delta J$ values for the probe transitions.  Combined
with the assignment of the pump transition, this allows the final state term value, symmetry, and total angular
momentum quantum numbers to be determined for the terminal state of each observed probe transition.  Even
in the homogeneously and strongly saturated case, polarization effects remain for low to modest $J$ values
due to the fact that not all possible $M$ values can be pumped.

\section{Acknowledgements}
Kevin K. Lehmann recognizes Steven L.Coy for conversations on this topic over many years and his and Aleksandra 
Foltynowicz's encouragement to publish these results.  He also acknowledges support for the US National Science Foundation.

\newpage

 \bibliography{Polarization_Ratios.bib}
 
 \newpage

\begin{figure}[t]
\begin{center}
\centering\includegraphics[width=20cm]{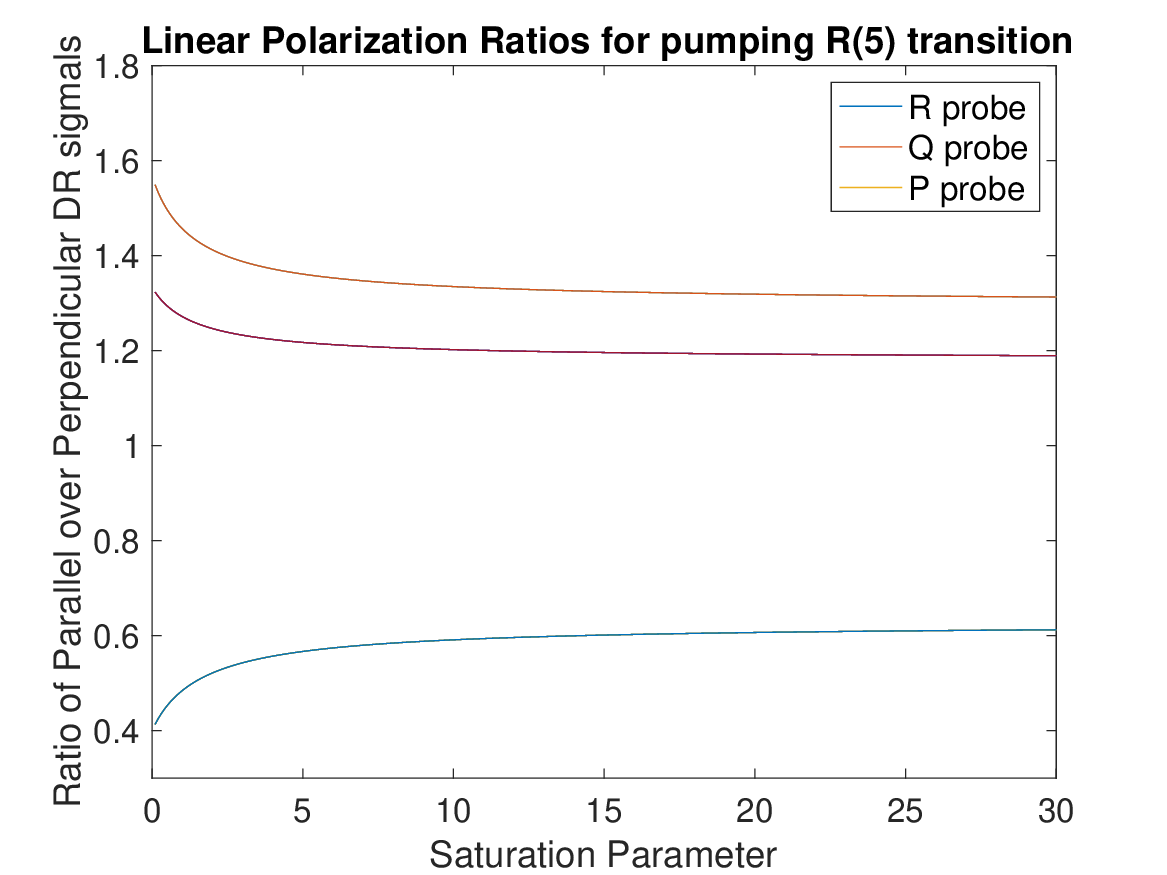}  
\caption{Probe linear polarization ratios as a function of saturation parameter for R(5) pump transition }
\label{linear_R(5)_pump}
\end{center}
\end{figure}
\newpage

\begin{figure}[t]
\begin{center}
\centering\includegraphics[width=20cm]{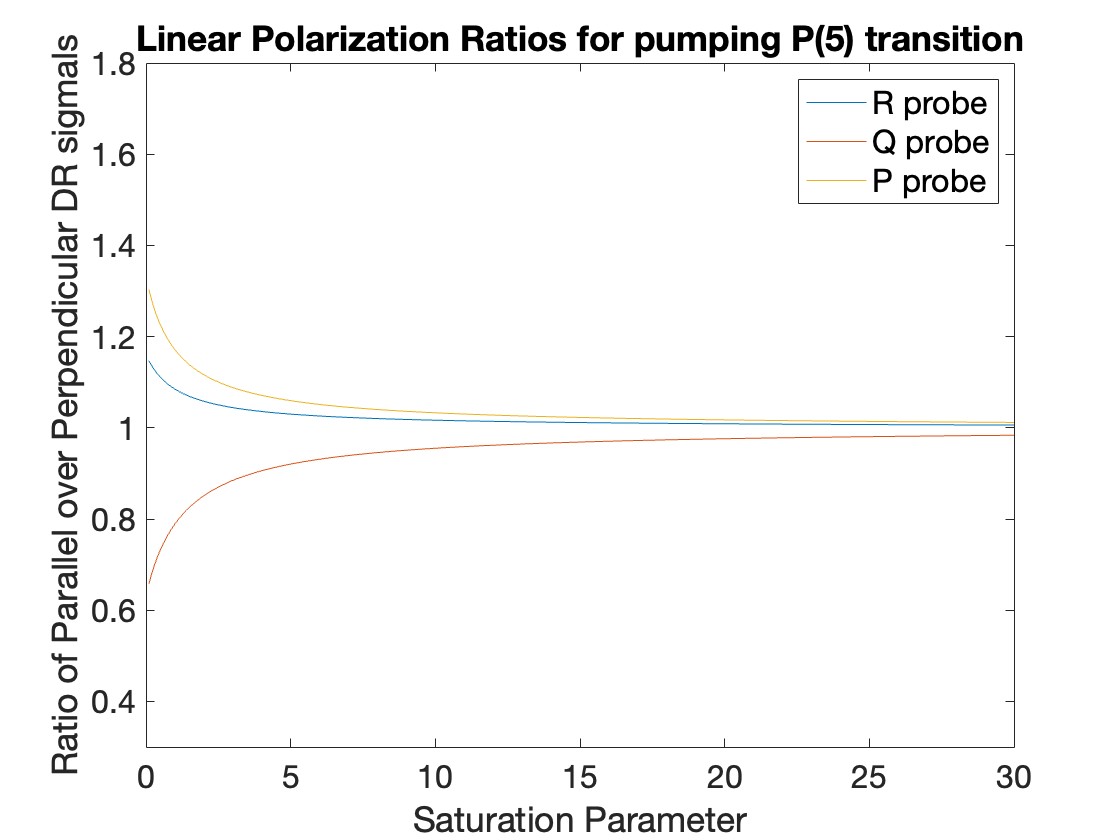}  
\caption{Probe linear polarization ratios as a function of saturation parameter for P(5) pump transition }
\label{linear_P(5)_pump}
\end{center}
\end{figure}
\newpage

\begin{figure}[t]
\begin{center}
\centering\includegraphics[width=20cm]{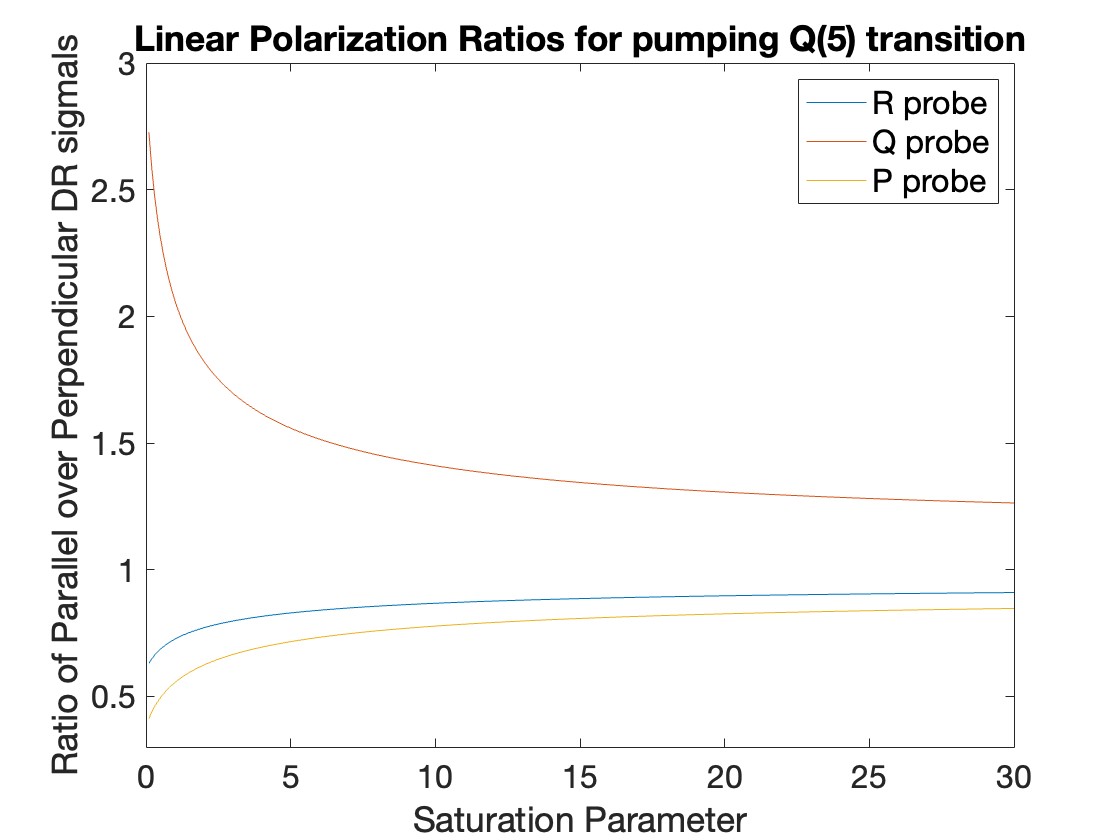}  
\caption{Probe linear polarization ratios as a function of saturation parameter for Q(5) pump transition }
\label{linear_Q(5)_pump}
\end{center}
\end{figure}
\newpage

\begin{figure}[t]
\begin{center}
\centering\includegraphics[width=20cm]{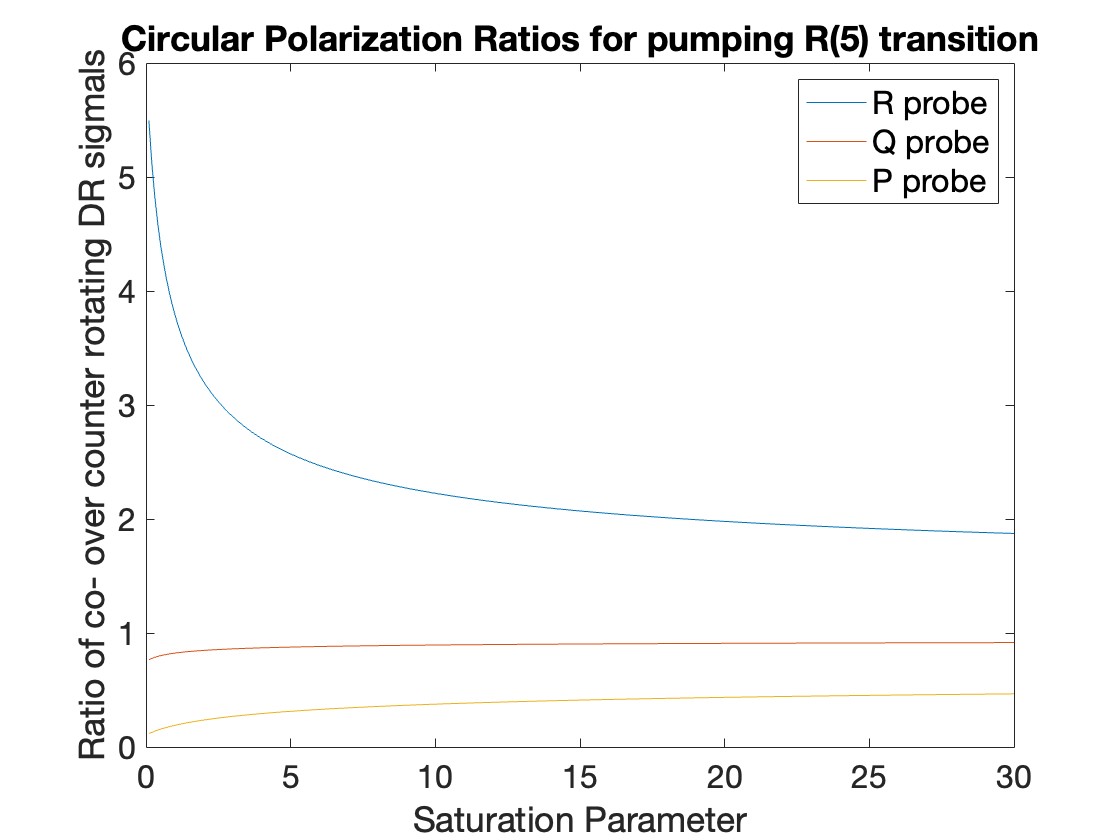}  
\caption{Probe circular polarization ratios as a function of saturation parameter for R(5) pump transition }
\label{circular_R(5)_pump}
\end{center}
\end{figure}
\newpage

\begin{figure}[t]
\begin{center}
\centering\includegraphics[width=20cm]{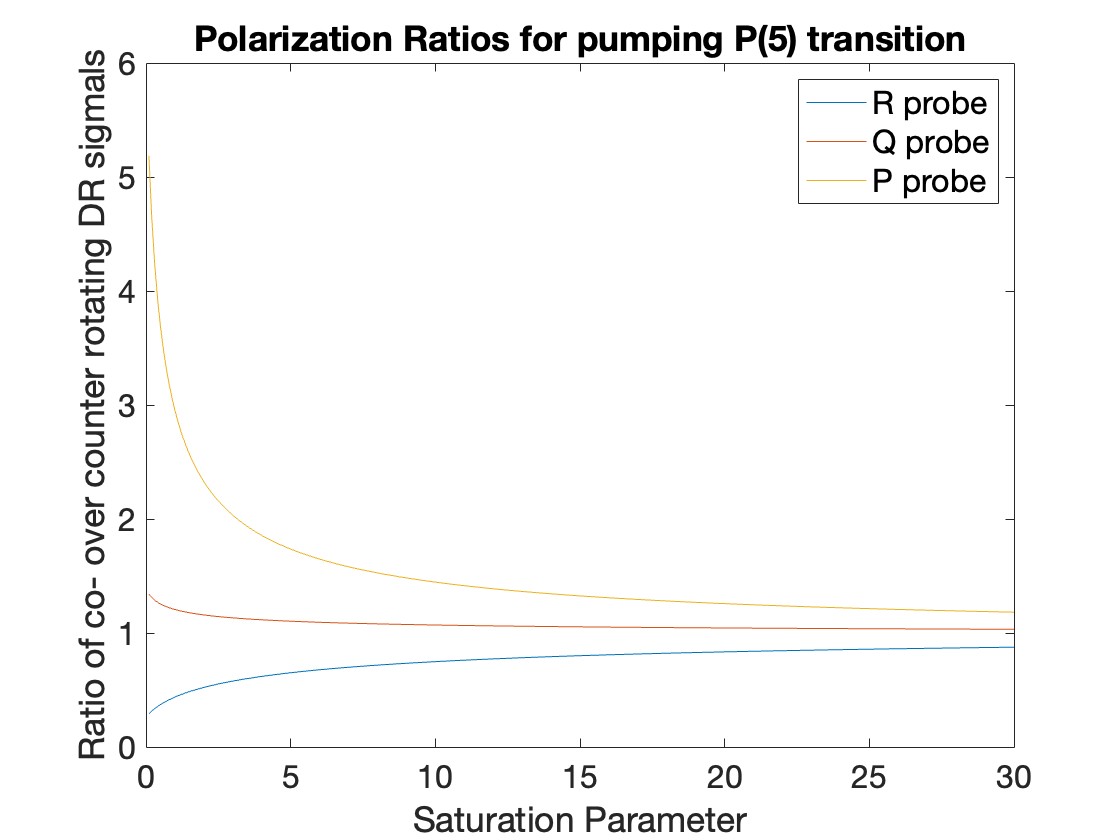}  
\caption{Probe circular polarization ratios as a function of saturation parameter for P(5) pump transition }
\label{circular_P(5)_pump}
\end{center}
\end{figure}
\newpage

\begin{figure}[t]
\begin{center}
\centering\includegraphics[width=20cm]{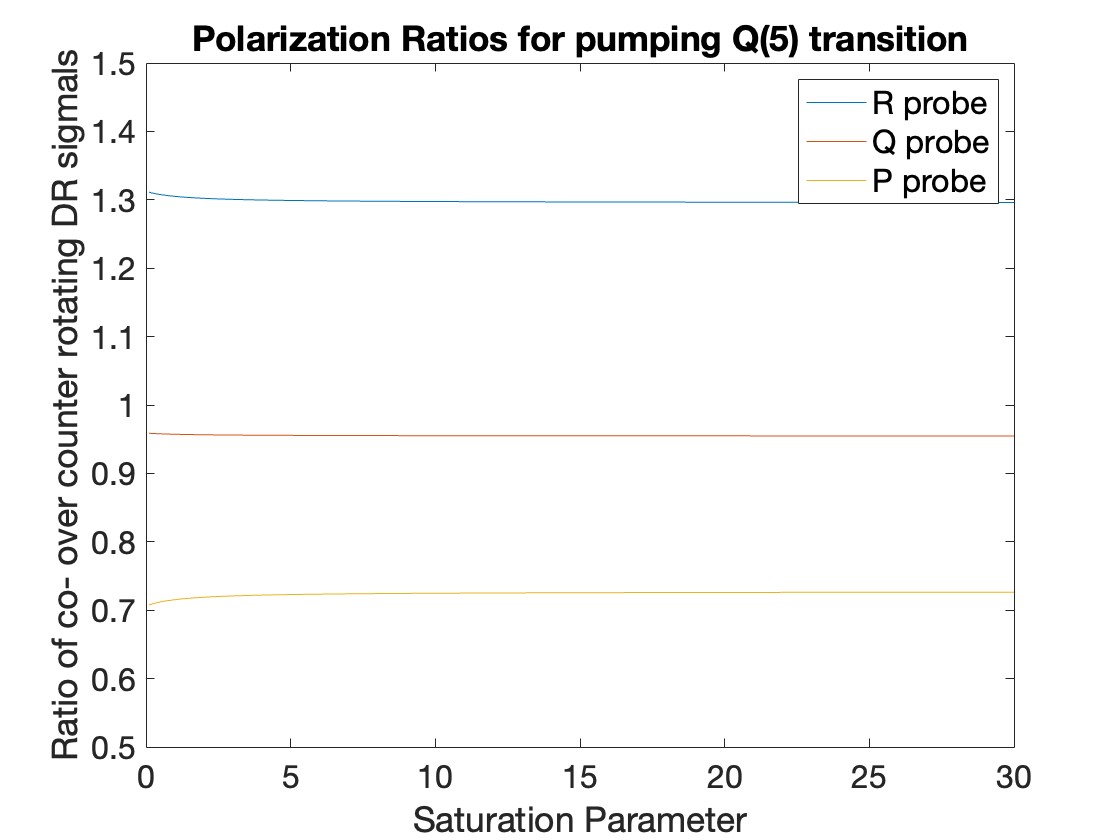}  
\caption{Probe circular polarization ratios as a function of saturation parameter for Q(5) pump transition }
\label{circular_Q(5)_pump}
\end{center}
\end{figure}
\newpage

\end{document}